\documentclass{article}
\usepackage{graphicx} 
\usepackage[utf8]{inputenc}
\usepackage{amsmath}
\usepackage{amssymb}
\usepackage{amsthm}
\usepackage{booktabs}
\usepackage{dsfont}
\usepackage[a-1b]{pdfx} 
\usepackage{hyperref}
\usepackage{csquotes}
\usepackage{pdflscape}
\usepackage[ruled]{algorithm2e}
\usepackage[a4paper, total={6in, 8.5in}]{geometry}
\hypersetup{colorlinks,linkcolor={black},citecolor={blue},urlcolor={black}}  
\usepackage[english]{babel}
\usepackage{comment}
\usepackage{bbm}
\usepackage{float}
\usepackage{multirow}
\usepackage{etoolbox}

\usepackage[round]{natbib}
\usepackage[toc,page]{appendix}

\usepackage{authblk}

\newtheoremstyle{mystyle}
  {\topsep}
  {\topsep}
  {\normalfont}
  {}
  {\bfseries}
  {.}
  {.5em}
  {}

\theoremstyle{mystyle}

\title{Conformal inference for cell type annotation with graph-structured constraints}
\author[1]{Daniela Corbetta}
\author[1]{Livio Finos}
\author[2]{Ludwig Geistlinger}
\author[1]{Davide Risso}
\affil[1]{Department of Statistical Sciences, University of Padova}
\affil[2]{Center for Computational Biomedicine, Harvard Medical School}
\date{}

\begin{document}
\date{\today}
\maketitle

\begin{abstract}
Conformal prediction is a framework for constructing prediction sets for machine learning models, 
relying solely on the exchangeability of training and test data and without requiring to specify a parametric distribution. Despite its wide applicability and popularity, its application in single-cell transcriptomics remains underexplored. This paper addresses this gap by developing an approach that leverages the rich information about cell-type relations, encoded in the graph structure of cell ontologies, to enhance the interpretability of reference-based cell-type annotation. Leveraging conformal risk control, we develop a novel conformal algorithm for graph-structured predictions and  we demonstrate how incorporating graph constraints can improve the interpretation of cell-type predictions. This approach aims to generate more coherent conformal sets that align with the inherent relationships among classes, facilitating clearer and more intuitive interpretations of model predictions. Additionally, we provide a technique to address non-exchangeability, particularly when the cell-type distribution changes between training and test datasets. We implemented our method in the open-source R package \textit{scConform}, available at \url{https://bioconductor.posit.co/packages/release/bioc/html/scConform.html}.
\end{abstract}

\section{Introduction}
\label{s:intro}

The advent of single-cell RNA sequencing technology has opened new avenues to unravel the intricate cellular landscape of heterogeneous tissues \citep{eberwine2014promise}. 
The rapid advancement in single-cell technologies, both sequencing-based and imaging-based \citep[e.g.,][]{chen2015spatially}, has led to the generation of diverse datasets, providing valuable insights into cellular heterogeneity, gene expression patterns, and the complex interactions between different cell types within a tissue.

Despite its promise, interpreting single-cell data remains challenging, particularly in identifying distinct cell types within complex cellular ecosystems. 
Traditional approaches to cell type prediction can be classified in \textit{reference-based} and \textit{marker-based} approaches: the former involve selecting an already annotated dataset from a similar sample as a reference, training a classification model on it, and then using the fitted model to predict the cell types in the new dataset; the latter involve a clustering step followed by either manual or automatic identification of cell types based on the expression of a set of marker genes \citep{amezquita2020orchestrating}.

\begin{figure}
    \centering
\includegraphics[width=\linewidth]{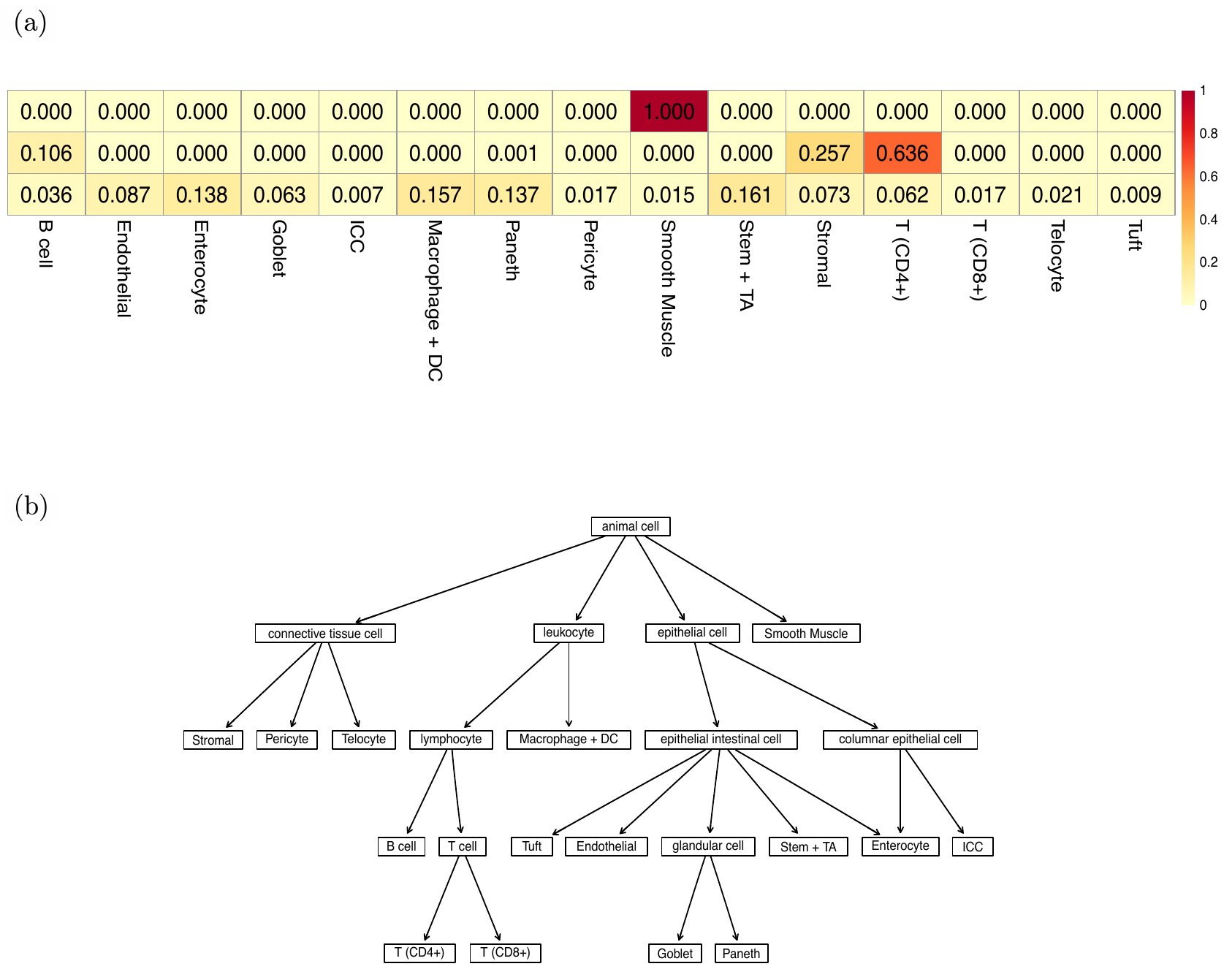}
    \caption{(a) Predicted probabilities for the 15 considered cell types for three different cells. (b) Graph-structure of the 15 cell types of the {\it Mouse Ileum} data, according to the Cell Ontology \citep{diehl2016cell}.}
    \label{fig:tree}
\end{figure}

The task of cell type prediction is crucial in many biological or clinical setting. A typical example is the case in which a laboratory receives data from a new patients, and needs to predict the cell types of the new cells based on the data that they have already available.
These approaches usually lack uncertainty quantification, potentially resulting in inaccurate biological interpretations and misguided subsequent research efforts. Even when the classifier provides such quantification, analysis workflows typically ignore it and simply assign the cell to the highest estimated cell-type probability.
For instance, consider a scenario where we have 15 different cell types. A classifier $\hat{f}(x)$ is trained on the reference dataset to predict cell types for the cells in a query dataset. The classifier provides estimated probabilities for each cell type, with the final prediction being the label associated with the highest probability. This approach can result in varying levels of confidence across predictions. For example, panel (a) of Fig. \ref{fig:tree} illustrates three different scenarios:
\begin{itemize}
    \item For the first cell (first row of the table), the model is highly confident, providing an estimated probability of 1 for a particular cell type (\textit{Smooth Muscle}).
    \item For the second cell, there is some uncertainty, but the point prediction (\textit{T (CD4+)}) remains fairly reliable.
    \item For the third cell, the point prediction is almost meaningless due to the presence of several classes with comparable predicted probabilities (\textit{Enterocyte}, \textit{Macrophage}, \textit{Paneth}, \textit{Stem + TA}).
\end{itemize}

These scenarios highlight the different levels of information conveyed by point predictions. Relying on a single label in each case can be problematic, as it does not accurately reflect the underlying uncertainties or complexities.

Conformal prediction provides confidence intervals for predictions generated by any machine learning model, applicable to both regression and classification tasks \citep{vovk2005algorithmic}. This methodology is model-agnostic and offers finite-sample, frequentist theoretical guarantees that the prediction set contains the true label with high probability, making it straightforward and widely applicable across various domains. Conformal prediction is particularly useful for performing distribution-free uncertainty quantification, in the sense that it does not require to specify a parametric distribution for the data or any specific condition such as smoothness.

When applied to a machine learning model predicting classes (e.g.,
labels, topics), conformal prediction is referred to as conformal classification \citep{angelopoulos2021}. In this context, the confidence is expressed over a set of possible classes rather than over an interval of values. This approach has gained popularity due to its flexibility and robustness, as demonstrated by numerous studies \citep{makili2012active,papadopoulos2008inductive} and methodological advancements \citep{johansson2017model,papadopoulos2011regression,devetyarov2010prediction}. In particular, one of the strengths of conformal inference is the fact that it provides prediction sets that are valid independently of the performance of the classifier used. However, if a classifier has a particularly bad accuracy, then the sets returned by the procedure are very wide, thus non-informative.
Despite its broad application, the exploration of conformal prediction in single-cell studies remains limited \citep{khatri2022uncertainty, sun2024, lopez2025conformal}. 

An important piece of information, typically overlooked by current analysis workflows, is the inherent relationships among cells, which are encoded as graph-structured constraints available through the Cell Ontology \citep[][see panel (b) of Fig. \ref{fig:tree} for an example]{diehl2016cell}.  The directed acyclic graph (DAG) provided by the Cell Ontology serves as an excellent framework for our approach.

 In particular, leveraging the graph structure that results from the ontological relationships of cells in the conformal framework is a promising avenue \citep{angelopoulos2022conformal}. Graph-structured labels are common in various fields; for instance, in object detection for images, labels such as {\it cat} and {\it dog} share the common ancestor {\it animal}, whereas they do not share a common ancestor with {\it car}. Similar hierarchical structures can be obtained from the Cell Ontology, where for example CD4+ T-cells share the common ancestor \textit{T-cells} with CD8+ T-cells and the (more distant) common ancestor \textit{lymphocyte} with B-cells (Fig. \ref{fig:tree}(b)). 
In these cases, standard conformal classification methods can produce sets consisting of labels situated in different parts of the graph (e.g., T (CD4+), Stromal, B cell, see second row of Fig. \ref{fig:tree}(a)), resulting in large shortest-path distances within the graph, complicating result interpretation, and reducing their practical utility. 

Rather than producing unconstrained prediction sets, the hierarchical structure of the Cell Ontology naturally suggests an alternative way to quantify uncertainty. Instead of returning multiple labels scattered across the graph, one can move upward along the ontology and report an ancestor of the predicted class. The lower the confidence in the point prediction, the higher (and thus more general) the ancestor that is returned. In this way, the method outputs a single, semantically meaningful label that summarizes the prediction set while respecting the underlying biological hierarchy. Equivalently, the prediction set can be seen as composed of all the descendants of that ancestor that are leaf nodes in the ontology. As a byproduct, we will obtain sets whose elements have the shortest path distance among them.

This paper aims to leverage the rich information encoded in the graph structure of predicted classes to enhance the interpretability of conformal sets. By doing so, we seek to address the limitations of standard methods and provide clearer, more meaningful results. 

We illustrate our approach and demonstrate its potential through applications to two publicly available datasets. First, we will use spatial transcriptomics data available as part of the \textit{MerfishData Bioconductor} package \citep{merfish}. Then, we will emulate a more realistic setting by applying our method to single-cell RNA-seq COVID-19 patient data from \cite{stephenson2021single}.

\section{Conformal inference}
\label{sec:confinf}
\subsection{Split conformal inference} \label{sec:split_conf}
Let $\{(X_i, Y_i)\}_{i=1}^m$ be a set of independent and identically distributed variables, where $X_i \in \mathbb{R}^p$ is a $p$-dimensional vector of explanatory variables, while $Y_i$ is a categorical response variable with $K$ classes, denoted as $1,\dots,K$. Assume that for a subset of $m$ observations $\{(X_i, Y_i)\}_{i=1}^m$ the values of $Y_i$ are known. Split conformal inference involves splitting the labelled data into two sets: a calibration set, $\{(X_i, Y_i)\}_{i=1}^{n}$, and a training set, $\{(X_i,Y_i)\}_{i=n+1}^m$. The training set will be used to build a classification model $\hat{f}$, that outputs estimated probabilities for each class, $\hat{f}(x) \in [0,1]^K$. For our application to cell type prediction, we assume that the $K$ classes correspond to the set of leaf nodes of the cell ontology and that the classifier $\hat{f}$ operates exclusively on them.
Given $\hat{f}$ and the calibration data, the goal is to construct a prediction set $C(X_{new}) \subseteq \{1,\dots,K\}$ for a new unlabelled observation $X_{new}$ that satisfies

\begin{equation}
    \label{eq:coverage}
    P\left(Y_{new}\in C(X_{new})\right) \geq 1-\alpha
\end{equation}

for a user-chosen error rate $\alpha$. The probability statement in Equation \eqref{eq:coverage} is calculated over the random draws of both the calibration data, used to construct the set $C$, and the new test point $(X_{new}, Y_{new})$. A prediction set is said to be valid if Equation \eqref{eq:coverage} is satisfied.
Conformal inference is distribution-free and provides finite-sample validity, assuming that the calibration data are exchangeable with the new data.

The split conformal inference algorithm proceeds as follows \citep{papadopoulos2002inductive}:
\begin{enumerate}
    \item For the data in the calibration set, $\{(X_i,Y_i)\}_{i=1}^n$, obtain the \textit{conformal scores}, $s_i=1-\hat{f}(X_i)_{Y_i}, \;i=1,\dots,n$. These scores will be high when the model is assigning a small probability to the true class, and low otherwise.
    \item Obtain $\hat{q}$, the $\lceil(1-\alpha)(n+1)\rceil/n$ empirical quantile of the conformal scores.
    \item Finally, for a new observation $X_{new}$, construct a prediction set by including all the classes for which the estimated probability is higher than $1-\hat{q}$: $C(X_{new})=\{y: \hat{f}(X_{new})_y\geq 1-\hat{q}\}$. 
\end{enumerate}

\subsection{Conformal risk control}\label{sec:crc}

Conformal risk control \citep{angelopoulos2022conformal} 
generalizes conformal inference to settings in which miscoverage, i.e. the probability that the prediction set fails to contain the true value, is not the only way to express reliability of the results. Notable examples include multi-label classification, scenarios where misclassifying certain labels has higher costs, or hierarchical label structures.

In this framework, two main components are required:
\begin{enumerate}
    \item A loss function to control, ensuring its expected value remains below a pre-specified threshold;
    \item an algorithm to build prediction sets that depends on a parameter $\lambda$. 
\end{enumerate}
Any algorithm is adequate as long as prediction sets are nested, with larger $\lambda$ resulting in broader sets. Moreover, the loss function needs to be monotone and non-increasing in $\lambda$ (i.e. lower risk with bigger sets).

In formulas, let $L_i(\lambda) \in (-\infty, B]$ be the loss function for the $i-th$ observation  and the prediction set obtained at $\lambda$. Suppose, as in the case of split conformal inference, that we have $n$ labelled observations, $\{(X_i,Y_i)\}_{i=1}^n$, which constitute the calibration set. The aim is to build a prediction set for a new test point $X_{new}$ by choosing $\lambda$ to satisfy
\begin{equation}
\label{eq:risk}
    E[L_{new}(\hat{\lambda})]\leq \alpha.
\end{equation}

\cite{angelopoulos2022conformal} proved that selecting $\hat{\lambda}$ as
\begin{equation}
\label{eq:lambda}
    \hat{\lambda}=\inf\left\{\lambda:\hat{R}_n(\lambda) \leq \alpha -\frac{B-\alpha}{n}\right\},
\end{equation}

where $\hat{R}_n=(L_1(\lambda)+\dots+L_n(\lambda))/n$ is the empirical risk over the observations in the calibration set, satisfies the previous inequality. 

\subsection{Conformal risk control for graph-structured labels}\label{sec:graph_proc}
We develop our approach as a special case of the broader framework of conformal risk control \citep{angelopoulos2022conformal}. 
Let $\hat{y}(x)$ be the class with maximum estimated probability returned by the classification model $\hat{f}$. Given a directed acyclic graph, 
define $\mathcal{P}(v)$ as the set of descendant nodes and $\mathcal{A}(v)$ as the set of ancestor nodes of $v$.
Let $N$ denote the set of leaf nodes of the graph (i.e. nodes with no outgoing edges). Define $\mathcal{L}(v)$ as the set of leaf nodes that are descendants of the node $v$, i.e. $\mathcal{L}(v)=\mathcal{P}(v) \cap N$. For each node $v$, define a score $g(v,x)$ as the sum of the predicted probabilities of the leaf nodes that are descendants of $v$: $g(v,x)=\sum_{i \in \mathcal{L}(v)} \hat{f}(x)_i$.

Our proposal to construct the prediction sets $C_\lambda(x)$ builds directly on these scores. Among all ancestors of the predicted class $\hat{y}(x)$, we identify the ancestor $v$ whose score is the smallest among those exceeding the threshold $\lambda$. Intuitively, this node represents the most specific ancestor of $\hat{y}(x)$ whose cumulative probability mass is at least $\lambda$. All leaf descendants of this ancestor are included in the prediction set.

This operation alone, however, is insufficient within the conformal risk control framework. In the presence of ramifications in the DAG (i.e. multiple parents for a node), restricting the prediction set to the leaf descendants of this single ancestor may lead to prediction sets that are not nested as $\lambda$ increases, and consequently to a loss function that is not monotonic in $\lambda$. To ensure the required nesting property, we additionally include the leaf descendants of all ancestors of $\hat{y}(x)$ whose scores are strictly below $\lambda$. This construction guarantees that the prediction sets expand monotonically as a function of $\lambda$, as required by conformal risk control.

In formulas, the prediction sets are built as follows:

\[
C_\lambda(x)
=
\bigcup_{\substack{
a \in \mathcal A(\hat y(x)) \\
g(a,x) \le \lambda
}}
\mathcal L(a)
\;\cup\;
\mathcal L(v),
\]
where $v \in \mathcal A(\hat y(x))$ is defined as
\[
v
=
\arg\min_{\substack{
u \in \mathcal A(\hat y(x)) \\
g(u,x) \ge \lambda
}}
g(u,x).
\]
In the case of ties, we select among the minimizers the ancestor closest to $\hat y(x)$ in terms of shortest path distance.

The loss function can be any monotone distance between the sets and the true class, based on shortest paths or other graph-based distances. However, for interpretability and comparability with previous results, we apply the miscoverage loss:
\begin{equation*}
    L_i(\lambda)= \begin{cases}
            1 & \text{if } y_i \notin C_\lambda(x_i)\\
            0 & \text{if } y_i \in C_\lambda(x_i)
        \end{cases}
\end{equation*}

When $\lambda=\hat{\lambda}$ as defined by Equation \eqref{eq:lambda}, this choice still guarantees that
$P(Y_{new}\notin C_{\hat{\lambda}} (X_{new})) \leq \alpha$.

\begin{figure}[H]
    \centering
    \includegraphics[width=0.7\linewidth]{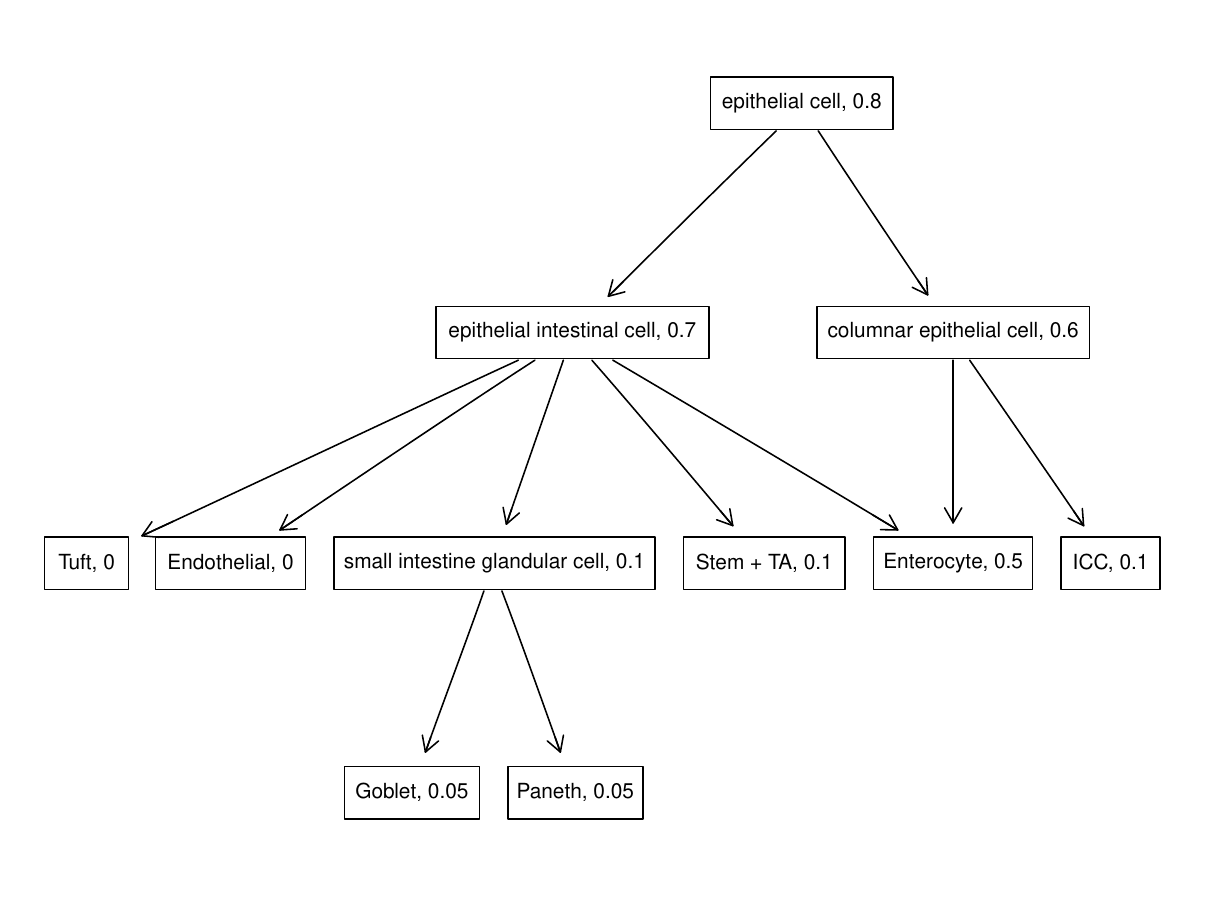}
    \caption{Reduced ontology with scores $g(v,x).$}
    \label{fig:subOnto}
\end{figure}

As a clarifying example, consider a small DAG, derived from the Cell Ontology, shown in Fig. \ref{fig:subOnto}. The previously defined scores $g(v,x)$ have been added to each node. 

    Suppose that the predicted class is \textit{Enterocyte}. For $\lambda=0.63$ the smallest subgraph with a score greater than $\lambda$ includes \textit{Epithelial intestinal cell} and its descendants. 
However, if 
$\lambda$ were 0.6, we would have included a different part of the graph, encompassing \textit{columnar epithelial cell} and its descendants. Thus, we need to include this subgraph in the final prediction set as well. Therefore, the final prediction set includes \textit{epithelial cell} and all its descendants.

\section{Distributional shift}\label{sec:dis_shift}
The fundamental assumption of conformal inference and conformal risk control is that the data in the calibration and test sets are exchangeable. However, this assumption often does not hold in real-world applications, leading to a distributional shift. Examples include time series data where patterns change over time, medical diagnosis data from different hospitals or time periods, evolving user behavior on online platforms or evolving language usage in natural language processing. In the field of genomics, this phenomenon is often referred to as \textit{batch effects}, which are widespread in single-cell data \citep{hicks2018missing}.

Assume $\{(X_i, Y_i)\}_{i=1}^n \stackrel{iid}{\sim} P$ for the data in the calibration set, and $\{(X_i, Y_i)\}_{i=m+1}^{l} \stackrel{iid}{\sim} Q$ for the test set. Distributional shifts can be broadly categorized into two main types: covariate shift and label shift. In the covariate shift model, it is assumed that the marginal distribution of the explanatory variables changes between the calibration and test sets, but the conditional distribution of the response variable given the explanatory variables remains the same. Formally, we have $P=p(x)p(y|x)$ and $Q=q(x)p(y|x)$. In the label shift model, it is instead assumed that the marginal distribution of the response variable changes between the calibration and test sets, but the conditional distribution of the explanatory variables given the response variable remains the same. Formally, we have $P=p(y)p(x|y)$ and $Q=q
(y)p(x|y)$. In the setting of conformal inference, the problems of obtaining valid prediction sets under the covariate shift and label shift models have been addressed by \cite{tibshirani2019conformal} and \cite{podkopaev2021distribution}, respectively.

In genomics, the typical workflow involves using a pre-existing annotated dataset as a reference for a new, non-annotated dataset. This means that the reference and test datasets are derived from different experiments and may originate from different tissues or populations. 
In this context, two major factors can lead to distributional shifts:

\begin{enumerate}
\item \textbf{Technical and Biological Variations}: Datasets may differ due to technical or biological factors. Technical variations arise from differences in measurement technologies, equipment, laboratory conditions, or protocols. Biological variations occur because data come from different spatial locations, organs, individuals, or species. These differences lead to the covariate shift model, where the distribution of explanatory variables (i.e., genes expression) changes. 
\item \textbf{Differences in Cell Type Composition}: The cell type composition in the reference dataset may differ from that in the test dataset. This issue leads to label shift.
\end{enumerate}

These problems often occur simultaneously. In a recent work, \cite{barber2023conformal} provide a method to deal with unknown violations of the exchangeability assumption, meaning that neither the covariate shift model nor the label shift model need to hold. However, their method requires choosing weights $w_i\in [0,1], \;i=1,\dots,n$ for the observations in the calibration set. Higher weights are assigned to observations that are more likely to have a distribution similar to that of the test set data. This implies that some prior knowledge regarding the relationship between calibration and test data is required; such prior knowledge is rarely available in our motivating application.

To address the distribution shift between calibration and test data within single-cell datasets, we propose a two-step approach:
\begin{itemize}
    \item The first step involves mitigating unwanted variability that leads to covariate shift. Recent advancements in bioinformatics have introduced effective methods to address this problem and integrate diverse datasets. Notable examples of such methods include SMAI \citep{ma2024principled}, Harmony \citep{korsunsky2019fast}, LIGER \citep{welch2019single} and MNN \citep{haghverdi2018batch}. This paper does not focus on this problem, and we assume that systematic differences have been accounted for with one of the above-mentioned approaches, ensuring $\tilde{p}(x|y)=\tilde{q}(x|y)$.

    \item The second step assumes that the only source of distributional shift is differences in cell type composition, thus adhering to the label shift model. To address this issue, we propose a solution based on a resampling strategy described in the next paragraph. This approach aims to equalize cell type distributions between the reference and test datasets, thereby enhancing the reliability of subsequent analyses.
\end{itemize}

\subsection{Resampling strategy}\label{sec:resampling}

The proposed strategy involves several steps. First, we build a predictive model $\hat{f}$
  using only the data in the training set. Next, we randomly split the data in the test set into two subsets, 
$S_1$
  and 
$S_2$. We then apply the fitted predictive model to the data in 
$S_1$
  and obtain the estimated probabilities for each class
$\hat{p}_{S_1}(Y=i)$, $i=1,\dots,K$. Subsequently, we resample the data in the calibration set according to the estimated probabilities and use this resampled set as a calibration set to obtain prediction sets for the data in 
$S_2$. Finally, we swap the roles of 
$S_1$  and $S_2$ and repeat the procedure. This strategy can be generalized to more folds or to a leave-one-out approach; however, our analysis suggests that when there are sufficient observations in the test set, the two-fold resampling strategy is sufficient, offering the lightest computational burden.

Importantly, the validity of this approach relies on the use of a predictively consistent model, so that the estimated probabilities $\hat{p}_{S_1}(Y=i|X)$ converge to the true conditional class probabilities as the sample size increases. 
Under this assumption, the resampling step faithfully approximates the underlying data-generating mechanism, and the resulting prediction sets enjoy asymptotic coverage guarantees.
 
Although this assumption may appear restrictive, it does not undermine the practical utility or applicability of our method. In fact, using a poorly performing classifier would typically yield prediction sets that include nearly all possible labels, making them essentially uninformative. For this reason, focusing on classifiers that are at least approximately consistent is both reasonable and aligned with the goal of producing meaningful prediction sets.

More formally, under the assumptions described above, the resampling strategy can be interpreted as an approximation of the conformal risk under the test distribution $Q$. In particular, under the label shift model, following the same arguments in \cite{angelopoulos2022conformal}, the risk appearing in Equation~\eqref{eq:risk} can be written as
\[
E_{\{(X_i, Y_i)\}_{i=1}^n\sim P, \,(X_{new}, Y_{new})\sim Q}[L_{new}(\lambda)]
=
E_{P}\!\left[
\frac{q(Y)}{p(Y)} \, L_{new}(\lambda)
\right],
\]
that is, as an importance-weighted expectation with respect to the calibration distribution $P$. The likelihood ratio $q(y)/p(y)$ is not known, but the proposed resampling step approximates this weighted risk by drawing calibration observations with probabilities proportional to the estimated class frequencies in the test data. When the classifier is predictively consistent, these estimated frequencies converge to the true label proportions under $Q$, and the resampled empirical risk converges to the target risk. This provides a principled motivation for the resampling procedure as a correction for label shift in the conformal risk control framework.

\section{Application to the Merfish dataset}
\label{sec:case1}
To illustrate our method and compare it to standard split conformal inference, we  utilise publicly available data from the \textit{MerfishData Bioconductor} package \citep{merfish}, which provides gene expression information of cells from the mouse ileum \citep{petukhov2022cell}. To obtain single cell data from the original Merfish data, cells have been segmented with Baysor, which optimizes 2D cell boundaries considering joint likelihood of transcriptional composition and cell morphology \citep{petukhov2022cell}. 
We chose this dataset as it is a relatively simple biological system that has been thoroughly studied and is composed of a limited number of well-understood cell types. Note that while these data arise from a spatial transcriptomics experiment, here we do not use the spatial information. However, our framework can work with spatially-informed cell-type prediction methods provided that they return a quantitative uncertainty score \citep{danaher2022insitutype}.
After preprocessing, there are 5136 cells and 241 genes in total. For each cell, a label indicating the cell type is provided. Cell annotation has been performed based on known marker genes.

There are 15 different cell types, listed with their frequencies:
{\it B cell} (536), {\it Endothelial} (231), {\it Enterocyte} (1257), {\it Goblet} (299), {\it ICC} (31), {\it Macrophage + DC} (427), {\it Paneth} (328), {\it Pericyte} (102), {\it Smooth Muscle} (428), {\it Stem + TA} (580), {\it Stromal} (489), {\it T (CD4+)} (197), {\it T (CD8+)} (125), {\it Telocyte} (115), {\it Tuft} (18). Cells that were classified as {\it Myenteric Plexus} (31) or {\it Removed} (606) were excluded from the analysis.
The induced graph structure of these cell types was derived from the Cell Ontology \citep{diehl2016cell}, using the functionalities available in the \textit{ontoProc Bioconductor} package, and is visualized in Panel (b) of Fig. \ref{fig:tree}.

From this structure it should be clear that, for example, a conformal set that includes {\it T (CD4+)} and {\it Paneth} might be ambiguous for users, while a conformal set that respects the graph-structure provides a more intuitive interpretation. For instance, a conformal set made only by the cell types {\it T (CD4+)} and {\it T (CD8+)} can be interpreted as a {\it T cell} (i.e. their closest common ancestor).

Next, we evaluate the performance of our method and compare it with standard split conformal classification using these data. 
The dataset is randomly split in three subsets: a training set of $500$ observations, a calibration set of $1000$ observations, and a test set with the remaining $3663$ observations. Thanks to this random split, in this section we do not need to deal with label shifts.

For our predictive model, we employed a multinomial logit model fit on the training data.
We selected the top 50 genes with the highest biological variance in log-expression across the training dataset as predictive features \citep{scran}. 
The model achieved an estimated accuracy of 0.763 on the test set. However, it is important to emphasize that our primary focus is not on the model's predictive accuracy \textit{per se}. While it is likely that a state-of-the-art model will achieve better accuracy, we deliberately chose a simple model, whose scores are straightforward to interpret, to illustrate the benefits of returning prediction sets instead of point predictions. Any probabilistic model or machine learning method can potentially be used, provided it produces a quantitative score for each class alongside point predictions. 

The calibration dataset is now used to estimate the quantile to be used in the split conformal method (see Section \ref{sec:split_conf}) and the parameter $\lambda$ to be used in the graph-structured approach (see Section \ref{sec:graph_proc}), computed as in Equation \ref{eq:lambda}. In the analysis, we set $\alpha=0.1$.
We compare the two approaches based on three metrics: empirical coverage, the size of prediction sets, and the homogeneity of elements within these sets, measured as the average of the shortest path among included cell types.

The empirical coverage is adequate for both methods (0.910 for split conformal and 0.912 for graph-based sets).

\begin{figure}
    \centering
    \includegraphics[width=0.8\linewidth]{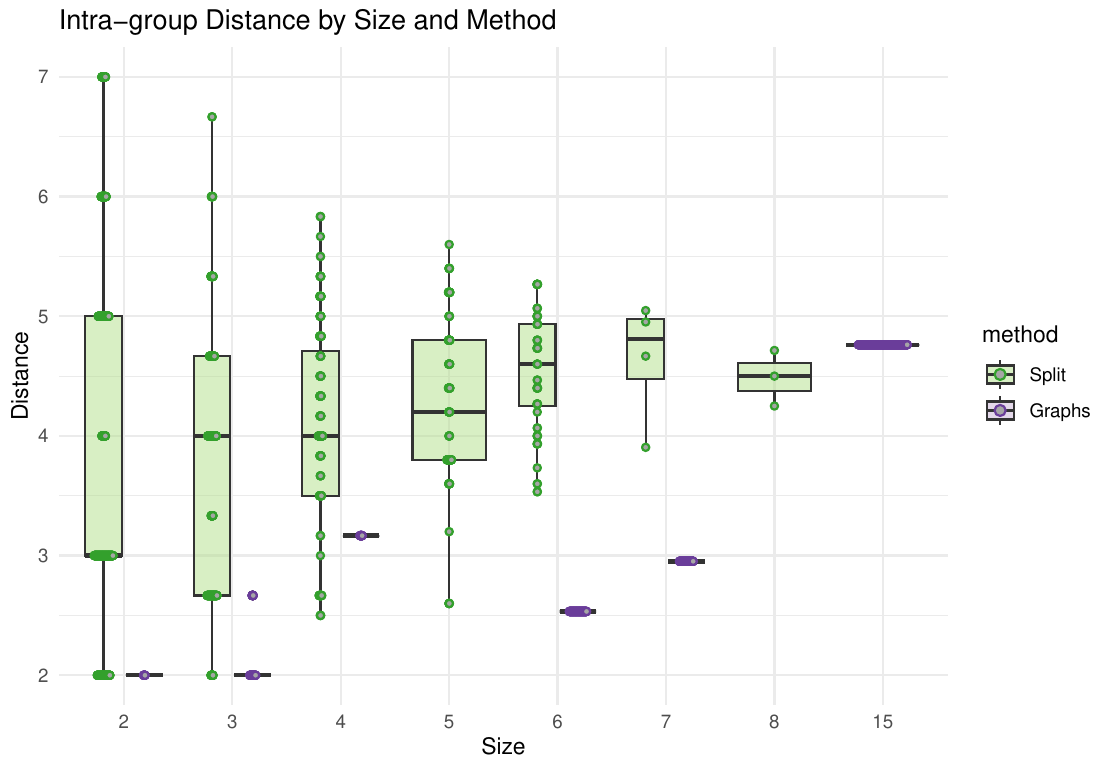}
    \caption{Distances among classes stratified by size and method}
    \label{fig:dist_bysize}
\end{figure}

The conformal sets produced by the graph-structured procedure are, on average, larger (average size 3.938 versus 1.863).  This is expected since the algorithm prioritizes labels closely related to the predicted class. 
Consequently, when the model assigns high probabilities to labels distant from the predicted class in the ontology, the algorithm must ascend higher in the ontology to find a subgraph with sufficient probability mass exceeding $\hat{\lambda}$.
However, labels within the ontology-based sets are, on average, closer to each other in terms of shortest path (average distance 1.148 versus 1.616 for split conformal sets), aligning with the objectives of this research.
Moreover, Fig.~\ref{fig:dist_bysize} shows the intra-group distances stratified by prediction set size and method. Across nearly all set sizes, the graph-based procedure yields sets with consistently lower intra-set distances compared to the split conformal method. This indicates that, regardless of the size of the prediction set, the labels selected by the graph method are more semantically related, reinforcing its ability to produce coherent prediction sets aligned with the ontology structure. 

\begin{figure}
    \centering
\includegraphics[width=0.7\linewidth]{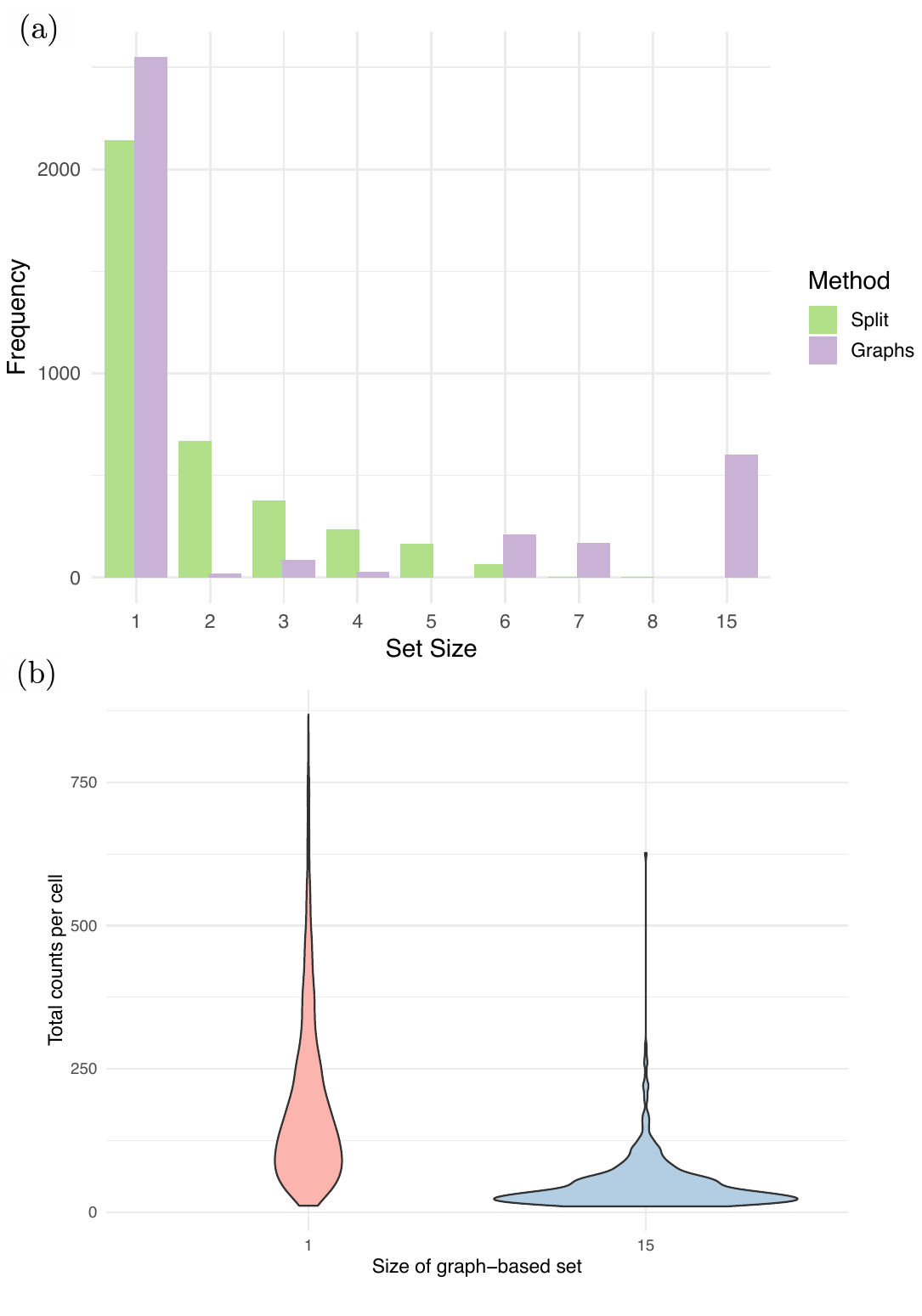}
    \caption{(a) Average Size of Conformal set for Standard Split Conformal and Graph-structured Conformal method. The spikes in 6 and 7 for the graph-based sets are caused by the structure of the considered ontology; (b) Distribution of the total counts per cell for cells whose corresponding graph-based prediction set contains only one label and for cells whose corresponding graph-based prediction set contains all 15 labels.}
    \label{fig:set_sizes}
\end{figure}

It is interesting to further explore the (average) number of cell types: Fig. \ref{fig:set_sizes} (a) compares the empirical distribution of set sizes among the two approaches. The graph-structured method more frequently produces sets containing a single label, thereby ensuring precise predictions of cell type. However, the average size is larger, as in some cases the prediction set comprises all 15 cell types in the study, leading to inconclusive predictions. From a practical point of view, it is preferable to return an inconclusive set rather than a set made of completely unrelated cell types, whose interpretation could be misleading. Analysts may further explore the set of cells whose prediction set comprises all cell types as they may be low-quality samples or cells of a cell type not included in the annotated reference.
Indeed, panel (b) of Fig. \ref{fig:set_sizes} highlights the differences in total counts per cell between cells with graph-based prediction sets containing only one label and those with prediction sets encompassing all 15 labels. The results suggest that cells with inconclusive prediction sets tend to be of lower quality compared to those with precise predictions.

To further show that having inconclusive sets might in some cases be informative, we obtained prediction sets for the two subsets of cells that were previously excluded from the analysis (\textit{Myenteric Plexus} and \textit{Removed}). 
Since these categories were not part of the reference set, it is desirable for the prediction sets to be inconclusive, reflecting the unreliability of assigning these cells to any of the reference cell types. Notably, the proportion of cells with graph-based prediction sets that encompass all the 15 cell types is 0.516 for \textit{Myenteric Plexus} cells and 0.691 for \textit{Removed} cells. In contrast, prediction sets obtained through split conformal inference are never inconclusive, incorrectly suggesting that these cells belong to one of the reference cell types.

\subsection{Choice of the calibration set size}
A key parameter to fix is the calibration set size $n$. Although split conformal prediction guarantees that
\[
P(Y_{\text{test}} \in C_{\text{test}}) \ge 1-\alpha
\]
in a frequentist sense for any choice of $n$, the random split induces variability in the empirical coverage. When $n$ is small, this variability can be substantial, resulting in a non-negligible probability of either undercoverage or overcoverage.

Selecting an appropriate $n$ therefore requires balancing several considerations: feasibility, accuracy, and computational cost.
\begin{itemize}
    \item \textbf{Feasibility.} The calibration set is drawn from the pool of already annotated data; the remaining data are used to train the classifier. The training set must remain sufficiently large to ensure good predictive performance.
    \item \textbf{Accuracy.} Larger calibration sets yield more stable coverage estimates, reducing the probability of deviating from the target level $1-\alpha$. Hence, $n$ should not be chosen too small.
    \item \textbf{Computational complexity.} Increasing $n$ also increases computational burden. Even when working with large annotated atlases containing hundreds of thousands of cells, it may be undesirable to use an excessively large calibration set.
\end{itemize}

In summary, the goal is to choose the smallest calibration set size that yields coverage sufficiently close to the desired level. For classical split conformal prediction, a helpful theoretical result is available \citep{vovk2012conditional}: the conditional coverage follows a Beta distribution \citep{vovk2012conditional},
\begin{equation}
\label{eq:covdist}
P\!\left( Y_{new} \in C(X_{new}) \,\middle|\, \{(X_i, Y_i)\}_{i=1}^n \right)
\sim \mathrm{Beta}(n+1-l,\, l),
\end{equation}
where $l = \lfloor (n+1)\alpha \rfloor$.  

\begin{figure}
    \centering
    \includegraphics[width=0.45\linewidth]{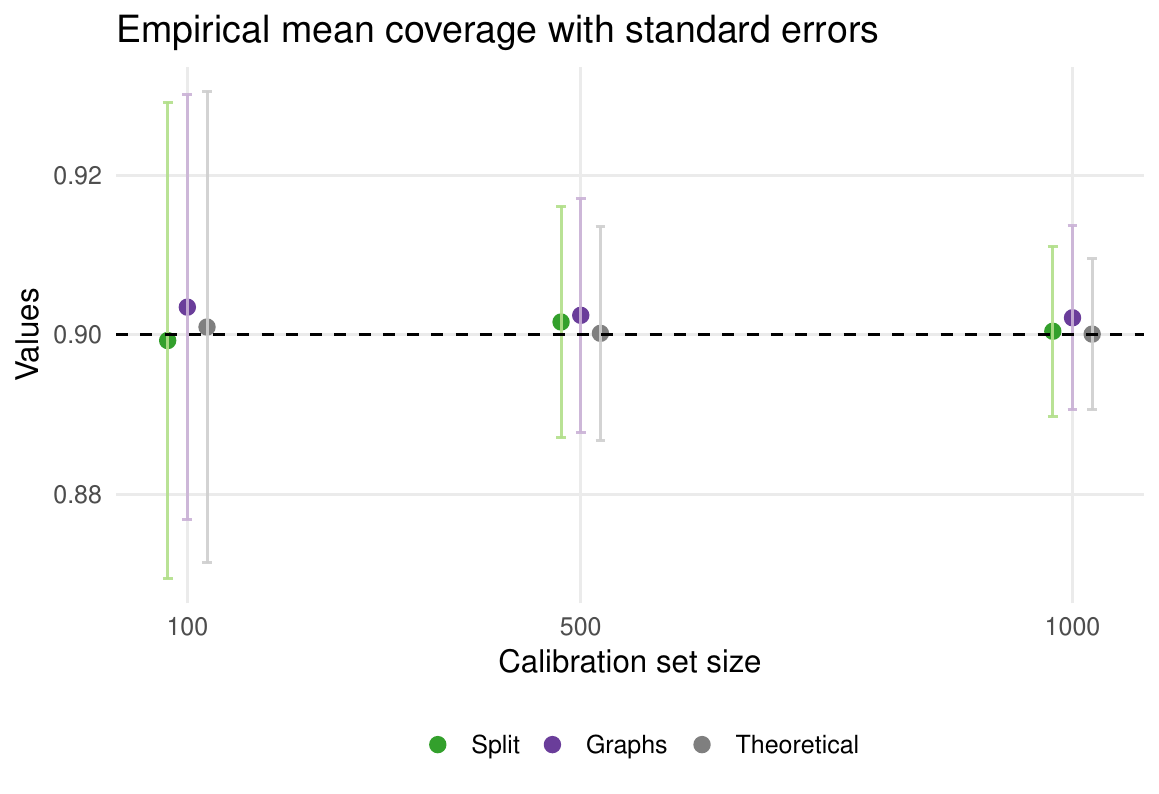}
    \includegraphics[width=0.45\linewidth]{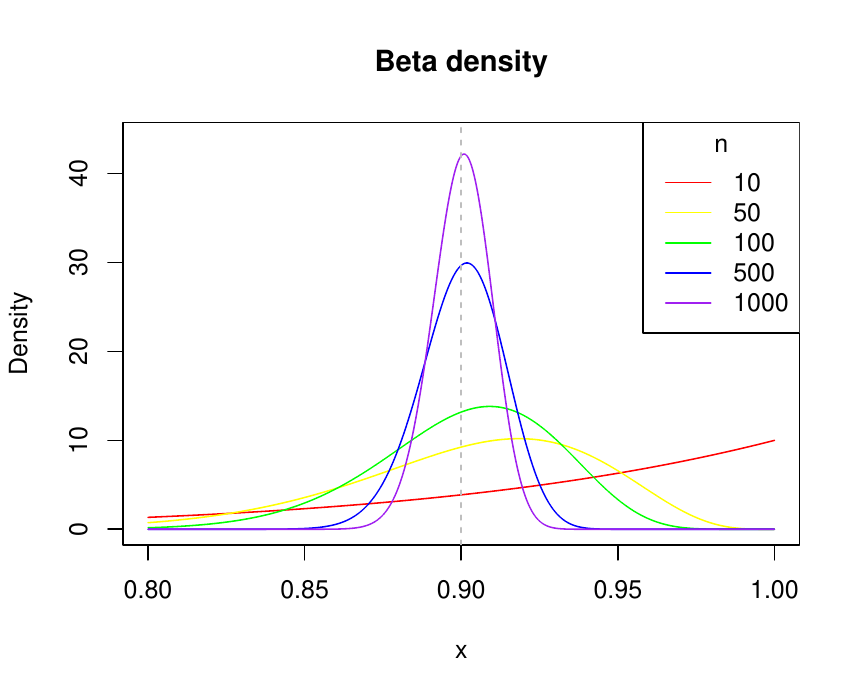}
    \caption{
(Left) Empirical mean coverage as a function of the calibration set size $n$ for split conformal prediction and the proposed graph-based method at $\alpha=0.1$. Error bars denote standard errors over 100 random splits; the theoretical standard error implied by the Beta distribution~\eqref{eq:covdist} is shown for reference.
(Right) Beta distributions of the conditional coverage for different values of $n$, illustrating how increasing the calibration size concentrates the coverage around the nominal level $1-\alpha=0.9$.
}

    \label{fig:caldist}
\end{figure}

For our graph-based method, no analogous closed-form result is available. Nevertheless, the left panel of Fig.~\ref{fig:caldist} compares the empirical mean coverages and standard errors of our algorithm with those of split conformal prediction, across 100 simulations and varying $n$ (with $\alpha = 0.1$). The standard errors implied by the theoretical Beta distribution~\eqref{eq:covdist} are also shown. Empirically, the variability of our method closely matches both standard conformal prediction and the theoretical benchmark.

Based on this observation, we recommend using the Beta distribution~\eqref{eq:covdist} as a guide for selecting $n$ (right panel of Fig. \ref{fig:caldist}). As a practical rule of thumb, when $\alpha = 0.1$, a calibration set size of $n = 1000$ provides excellent stability: approximately $96\%$ of the Beta density lies between $0.88$ and $0.92$, which is sufficiently close to the target level $0.9$.

\subsection{Dealing with label shift}
\label{sec:res}

In the previous section, we demonstrated the effectiveness of our strategy when the data in the calibration set and the test set are exchangeable. We will now address the common scenario where the label shift model applies, meaning the distribution of cell types differs between the reference and test sets. We will present two different examples.

In both examples, we use the multinomial-logit model trained in the previous section. We then sample the calibration and test data with different probabilities for each cell type. Under these conditions, there are no guarantees on the coverage of the sets, neither for the conformal nor for the graph-based procedure.
Since there are no guarantees on coverage, the procedure might either over-cover or under-cover. We provide an example for each case: in the first example, the empirical coverage of the prediction sets derived from the non-corrected procedure is lower than the nominal coverage, while in the second example, detailed in Appendix \ref{appendix-example}, it is higher. Our results suggest that the proposed strategy effectively addresses the problem in both cases, restoring the correct coverage.

\begin{figure}[H]
    \centering
    
\includegraphics[width=0.8\linewidth]{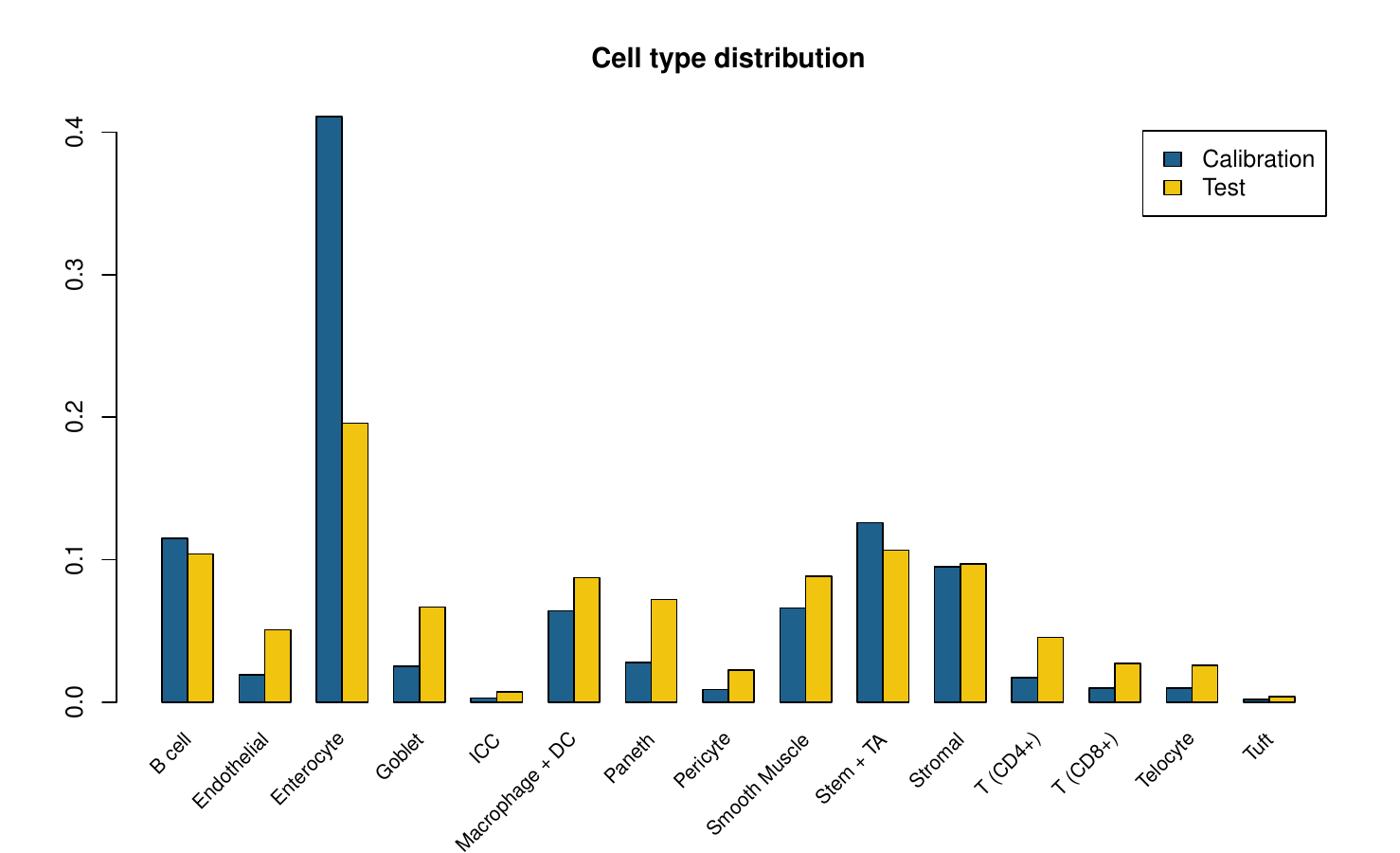}
    \caption{Distribution of cell types in calibration and test dataset.}
    \label{fig:dis_under}
\end{figure}

\begin{figure}[H]
    \centering
    
\includegraphics[width=\linewidth]{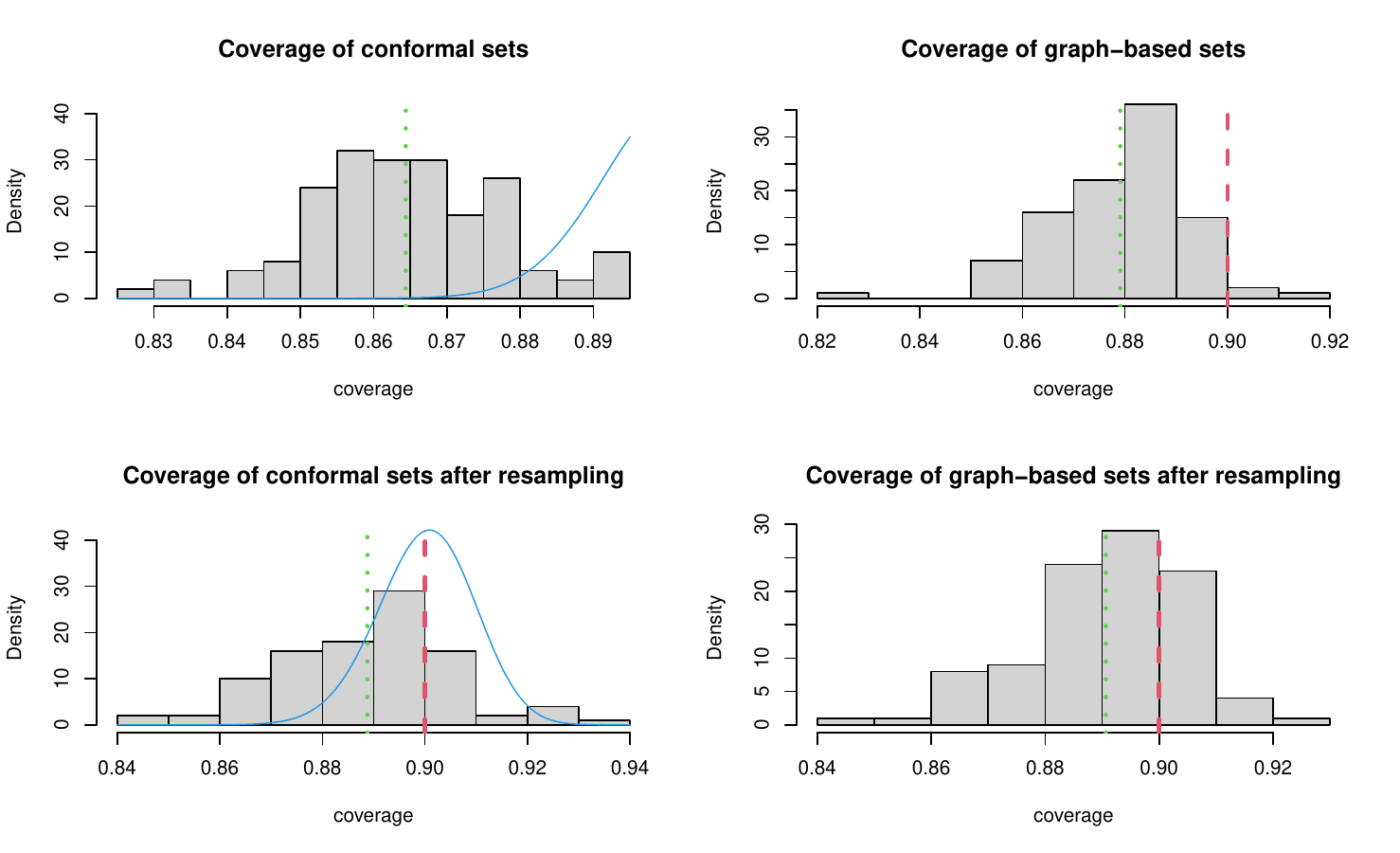}
    \caption{Empirical coverage of the prediction sets over 100 simulations obtained by conformal inference (first column) and by the graph-based procedure (second column) for the first case study. Results in the first row have been obtained without resampling, results in the second row have been obtained resampling. The red dashed lines correspond to the desired coverage, the green dotted ones to the empirical coverage.}
    \label{fig:umderEmpCov}
\end{figure}

Fig. \ref{fig:dis_under} illustrates the distribution of cell types in the calibration and test sets for the first example. The coverage of standard conformal inference is 0.878, and the coverage of the graph-based procedure is 0.885, both falling short of the nominal coverage of 0.9, indicating undercoverage. To ensure that this undercoverage is not merely a result of the chosen split, we repeated the procedure across 100 different splits of the calibration and test datasets, maintaining the same proportions of cell types. The mean empirical coverage across these 100 simulations is 0.864 (standard error, s.e. 0.0013) for the standard conformal method and 0.879 (s.e. 0.0012) for the graph-based procedure. The top row of Fig. \ref{fig:umderEmpCov} displays the distribution of coverage over these 100 simulations.

To further validate the results, we compared the empirical coverage distribution of the standard conformal method with its theoretical distribution \citep{vovk2012conditional},
overlaid on the histogram of empirical coverage in the top left panel of Fig. \ref{fig:umderEmpCov}, clearly showing a discrepancy between the empirical and theoretical distributions.

We then applied the resampling strategy discussed in Section \ref{sec:resampling} and repeated the analysis. For a single split, the empirical coverage was 0.907 for the conformal sets and 0.906 for the graph-based sets. The second row of Fig. \ref{fig:umderEmpCov} shows the distribution of empirical coverage over 100 simulations, with means of 0.888 (s.e. 0.0017) for the conformal sets and 0.891 (s.e. 0.0014) for the graph-based sets. These results demonstrate that the resampling strategy brings the empirical distribution closer to the theoretical one (bottom left panel). Although perfect alignment is unattainable due to sampling variability, this approach significantly improves coverage accuracy.

\section{Application to the COVID-19 dataset}
\label{sec:case2}
While Section \ref{sec:case1} was based on a real dataset, it demonstrated our framework in simplified scenarios: first, by randomly sampling training, calibration, and test sets from the same distribution, and second, by following the label shift model. We now apply our method in a more realistic setting using a subset of the dataset provided by \cite{stephenson2021single}. 

\begin{figure}
\centering
\includegraphics[width=0.8\linewidth]{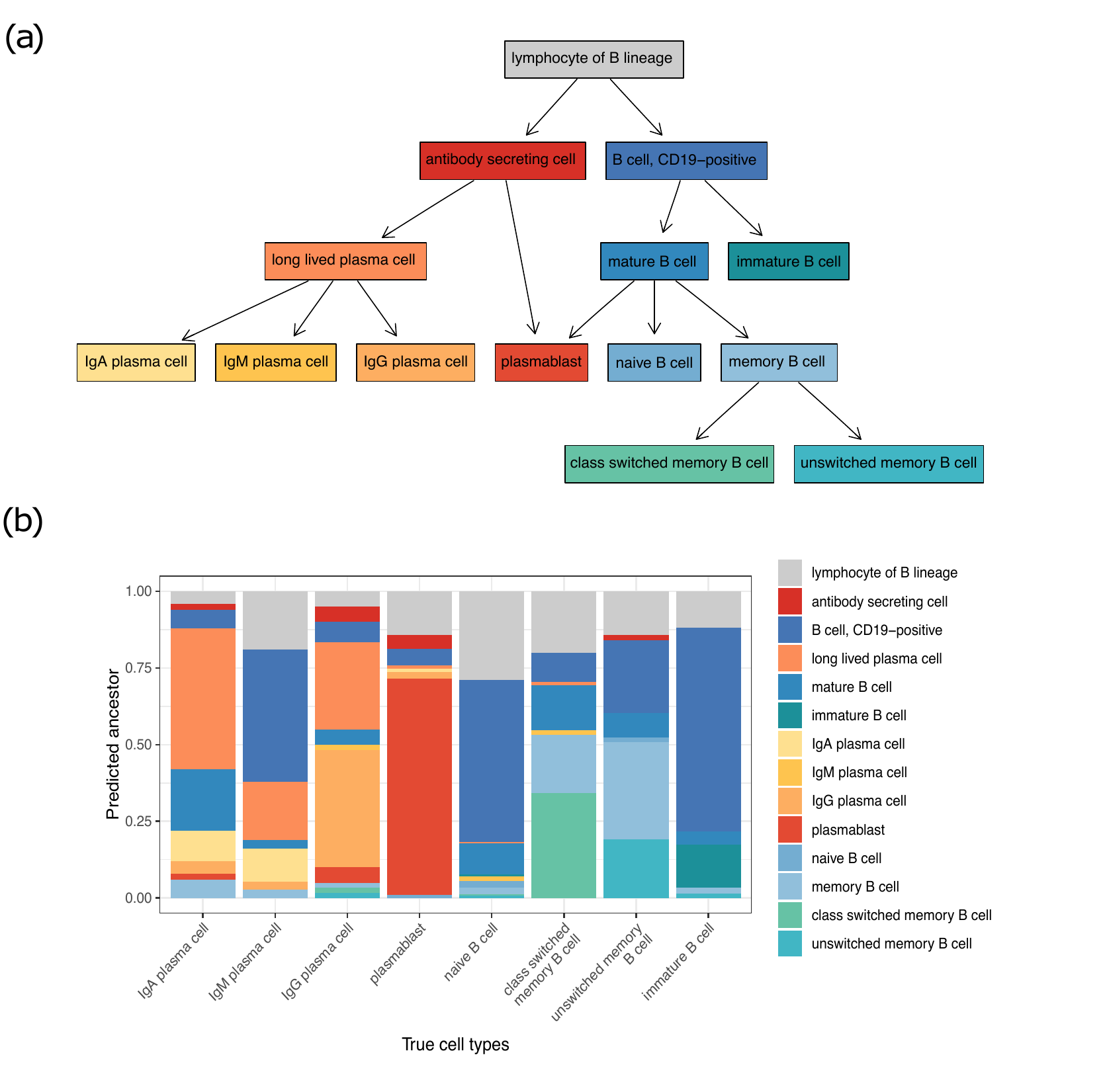}
    \caption{(a) Cell ontology for the subtypes of B cells considered in the Covid case study; (b) barplot of the frequencies of predictions assigned by our method to each true cell type, normalized so that proportions sum to 1 within each cell type. Cells have been coloured according to the colour scheme defined in (a).}
    \label{fig:covid}
\end{figure}

This dataset includes samples from both healthy donors and COVID-19 patients, ranging from asymptomatic to severely ill cases, collected across three medical centres in the United Kingdom. The original publication provided a comprehensive analysis of peripheral blood samples from a cross-sectional patient cohort. However, following the preprocessing described in \cite{Gilis2023}, we focused on eight distinct B cell subtypes, selected based on the availability of sufficient cell numbers per patient and an adequate number of patients for each disease category. These subtypes include naive B cells, immature B cells, class-switched memory B cells, unswitched memory B cells, IgG plasma cells, plasmablasts, IgA plasma cells, and IgM plasma cells. Panel (a) of Fig. \ref{fig:covid} illustrates the ontological relationships among these cell types. 
For simplicity, we restricted the analysis to data from the Newcastle centre. Additionally, we excluded samples obtained from healthy donors. After these preliminary steps, we obtained a dataset that comprised 34193 single cells from 43 different donors.

Our goal is to simulate the case where cell type predictions are needed for a new patient based on previously collected data, allowing us to assess the robustness of our method in a realistic clinical setting.
The dataset was therefore divided as follows: all cells from a single donor (patient ID: MH9143273, 1762 cells) were used as the test set, while the remaining cells constituted the reference dataset. However, the reference dataset was highly unbalanced, with naive B cells alone constituting 60\% of the cells. This imbalance could negatively impact model performance, leading to unreliable predictions and undermining the effectiveness of the resampling strategy. To address this, we downsampled the reference dataset to ensure a balanced representation of the different cell types. The final reference dataset contained 5616 cells. A random sample of 1000 of these cells was reserved as the calibration set, while the remaining 4616 cells were used as the training set. Cells in the calibration and in the training sets were annotated with a probabilistic model by \cite{grabski2022}.

Our graph-based procedures, both with and without the resampling correction, are compared to an oracle correction that resamples the calibration set according to the true observed frequencies of the test set. 

Non-resampled sets consistently exhibit empirical coverages higher than the nominal 0.9 (0.948 for conformal sets and 0.966 for graph-based sets). In contrast, 
the resampled procedures improve these results: with the oracle correction, empirical coverages reach 0.918 for conformal sets and 0.896 for graph-based sets, while with the estimated correction, they achieve 0.930 and 0.915, respectively. Remarkably, the results from the oracle and estimated corrections align closely.

By leveraging the ontology, our method consolidates multiple fine-grained labels into a single, biologically meaningful label for the prediction set. When the model is sufficiently confident, it outputs the corresponding leaf node; otherwise, it returns one of its ancestors. Panel (b) of Fig. \ref{fig:covid} illustrates the type of biological insight enabled by this approach.
Overall, our approach leads to biologically coherent and interpretable cell type labels, as underlined by the fact that the four cell types corresponding to the left-most leaf nodes of the ontology are mostly labelled as antibody secreting cells and its descendant (red and orange hues in Fig. \ref{fig:covid}). Conversely, the four right-most leaf nodes are mostly labelled as B cell descendants (blue and green hues).

Interestingly, our approach highlights inherent differences between cell types with respect to the classifier's performance. For instance, a large fraction of \textit{plasmablast} and \textit{IgG plasma cells} are correctly labelled as singletons, while most \textit{IgA plasma cells} and \textit{IgM plasma cells} are labelled as more generic cell types, sometimes incorrectly involving a different part of the ontology. Similar considerations apply to the right side of the ontology, where \textit{class-switched memory B cells} and \textit{unswitched memory B cells} are more easily classified to have a precise label compared to \textit{naive B cells} and \textit{immature B cells}, that are often classified more generally as \textit{B cell, CD19-positive}.

\section{Discussion}

This paper contributes to the exploration of graph-structured problems by incorporating graph-structured constraints into conformal prediction methods. Additionally, we introduce a simple and efficient strategy to address the label shift model, expanding the applicability of our method.

The idea of using conformal inference for cell type prediction in single-cell applications was first explored by \cite{khatri2022uncertainty}. However, their work is limited to standard conformal classification and does not consider the supplemental information encoded in the cell ontology. Our findings demonstrate that incorporating graph-structured constraints improves the interpretation of cell-type predictions in single-cell data, yielding prediction sets that are more coherent with the hierarchical relationships encoded in the ontology.

A central contribution of our approach is the explicit integration of uncertainty quantification with the semantic structure provided by the cell ontology. Rather than committing to a single fine-grained label, even in cases where the model is intrinsically uncertain, our method outputs structured prediction sets with finite-sample guarantees. This allows downstream analyses to explicitly incorporate model uncertainty and leaves open different choices for how to resolve or propagate this uncertainty depending on the scientific question. In other words, the method does not dictate a single post-processing strategy but instead provides reliable information that can be flexibly exploited. Developing principled ways to propagate these uncertainty-aware outputs through downstream analysis pipelines is an important direction for future research.

A second line of related work concerns hierarchical cell-type classification \citep[see e.g.][]{bernstein2021cello}. These methods aim to determine annotation granularity in a data-driven way, but lack formal uncertainty quantification. Consequently, they inherit the same issues as standard leaf-level classifiers: errors made at early stages of the hierarchy propagate downstream, and the user must rely on the quality of the trained hierarchical model, with no coverage guarantees. Our work is fundamentally different in this respect. By producing graph-constrained conformal prediction sets, we provide valid coverage at level $1-\alpha$ regardless of the underlying classifier’s accuracy. In practice, this means that the final label or set of labels reported by our approach is guaranteed to contain the correct cell type with the prescribed probability.

Beyond the specific construction detailed in our paper, there are multiple ways in which nested, graph-constrained prediction sets could be defined. While the method proposed in this work focuses on identifying the closest ancestor of the predicted class whose score reflects the model’s uncertainty, this is not the only possible construction that respects the directed acyclic graph structure of the ontology. For instance, one could define prediction sets by indexing ancestors according to their distance from the predicted class, including all leaves of ancestors up to a fixed number of steps in the hierarchy. Alternatively, one could rank all leaf nodes by their predicted probabilities, include leaves until a cumulative probability threshold is reached, and then return all leaves descending from the lowest common ancestor of the selected leaves. Both approaches yield nested prediction sets and satisfy the same conformal coverage guarantees. In our experiments on the MERFISH dataset, the distance-based construction systematically yielded larger prediction sets than our proposed method, resulting in reduced specificity. The probability-ranking construction instead led to prediction sets that were very similar to ours, with approximately 70\% of the prediction sets being exactly identical. For completeness and flexibility, we have implemented these alternative constructions in the accompanying R package, allowing users to select the strategy most appropriate for their application.

Despite these promising results, several challenges and limitations remain. Firstly, this work is restricted to acyclic graphs. Future research should aim to incorporate more complex graph structures into conformal prediction methods to enhance their applicability and robustness.

The COVID-19 case study also highlights an important direction for future work. While our current experiments do not focus on batch effects, understanding how distributional shifts affect coverage remains an open question. In particular, our resampling strategy is not designed to correct batch effects; rather, it is used solely to harmonize label frequencies between calibration and test sets, ensuring that distributional shift is limited to the specific scenario covered by the label shift model. Future research will investigate how batch effects interact with distributional shifts more broadly and will benchmark the most effective strategies to address them within conformal frameworks, with the final aim of identifying the optimal preliminary steps for applying our method and the best strategies for dataset integration.

Our framework works with any supervised or semi-supervised cell-type prediction algorithm that outputs a quantitative score for the uncertainty of the prediction, such as, but not limited to, probabilistic cell-type assignment \citep{zhang2019probabilistic}. Note that many of the most popular cell-type prediction algorithms, e.g., SingleR \citep{aran2019reference} and Azimuth \citep{hao2021integrated}, fall into this category.

Another key assumption in our work, necessary for the construction of the graph-based score $g(v, x)$, is that the reference cell types used for training and calibration must correspond exclusively to the leaf nodes of the cell ontology. This requirement ensures that the classifier $\hat{f}$ operates at the fine-grained resolution needed by the method. However, the methodology could be readily extended to accommodate reference datasets that contain cell-type annotations at intermediate granularities (i.e., non-leaf nodes). This extension is possible by leveraging the generality of the Conformal Risk Control framework. If a true label $Y_i$ corresponds to an internal, non-leaf node, the miscoverage loss function $L_i(\lambda)$ must be generalized to reflect the required coverage guarantee. Specifically, we can utilize the generalization suggested by the framework to define the loss as
\[
L_i(\lambda) = \mathbf{1}(\mathcal{L}(Y_i) \not\subseteq C_\lambda(X_i)),
\]
where $\mathcal{L}(Y_i)$ denotes the set of leaf descendants of the non-leaf label $Y_i$, and $C_\lambda(X_i)$ is the prediction set returned by our algorithm. This generalized loss ensures that miscoverage only occurs if the prediction set $C_\lambda(X_i)$ fails to fully encompass all the specific, fine-grained cell types that fall under the true, coarse label $Y_i$. Adapting the loss function in this manner would allow practitioners to seamlessly integrate reference data containing multi-granularity annotations during the calibration step.

\section*{Code availability}

Our method is implemented in the R package scConform, available at \url{https://github.com/ccb-hms/scConform}. The code to reproduce the analysis is available at \url{https://github.com/DanielaCorbetta/scGraphConformPaper}. A detailed tutorial on how to construct the cell ontology starting from the cell type labels is available at \url{https://bioconductor.posit.co/packages/release/bioc/vignettes/scConform/inst/doc/scConform.html}.

\section*{Data availability}
The mouse ileum Merfish data used in Section \ref{sec:case1} are publicly available from the \textit{MerfishData} Bioconductor package at \url{https://doi.org/10.18129/B9.bioc.MerfishData}. The preprocessed B cells COVID data used in Section \ref{sec:case2} are openly available in \texttt{Covid\_case/objects/sce\_Covid\_Bcells.rds}  at  \url{https://doi.org/10.5281/zenodo.10391097}.

\bibliography{mybiblio,reference}

\clearpage

\begin{appendices}

\begin{center}\Large\bf
    Appendix of ``Conformal inference for cell type annotation with graph-structured constraints''
\end{center}

\setcounter{figure}{0}  
\renewcommand{\figurename}{Supplementary Fig.}

\section{Additional example of label shift}
\label{appendix-example}

While in Section \ref{sec:res} we provide an example that addresses undercoverage, here we provide an example that deals with overcoverage.

\begin{figure}[H]
    \centering
    
\includegraphics[width=0.8\linewidth]{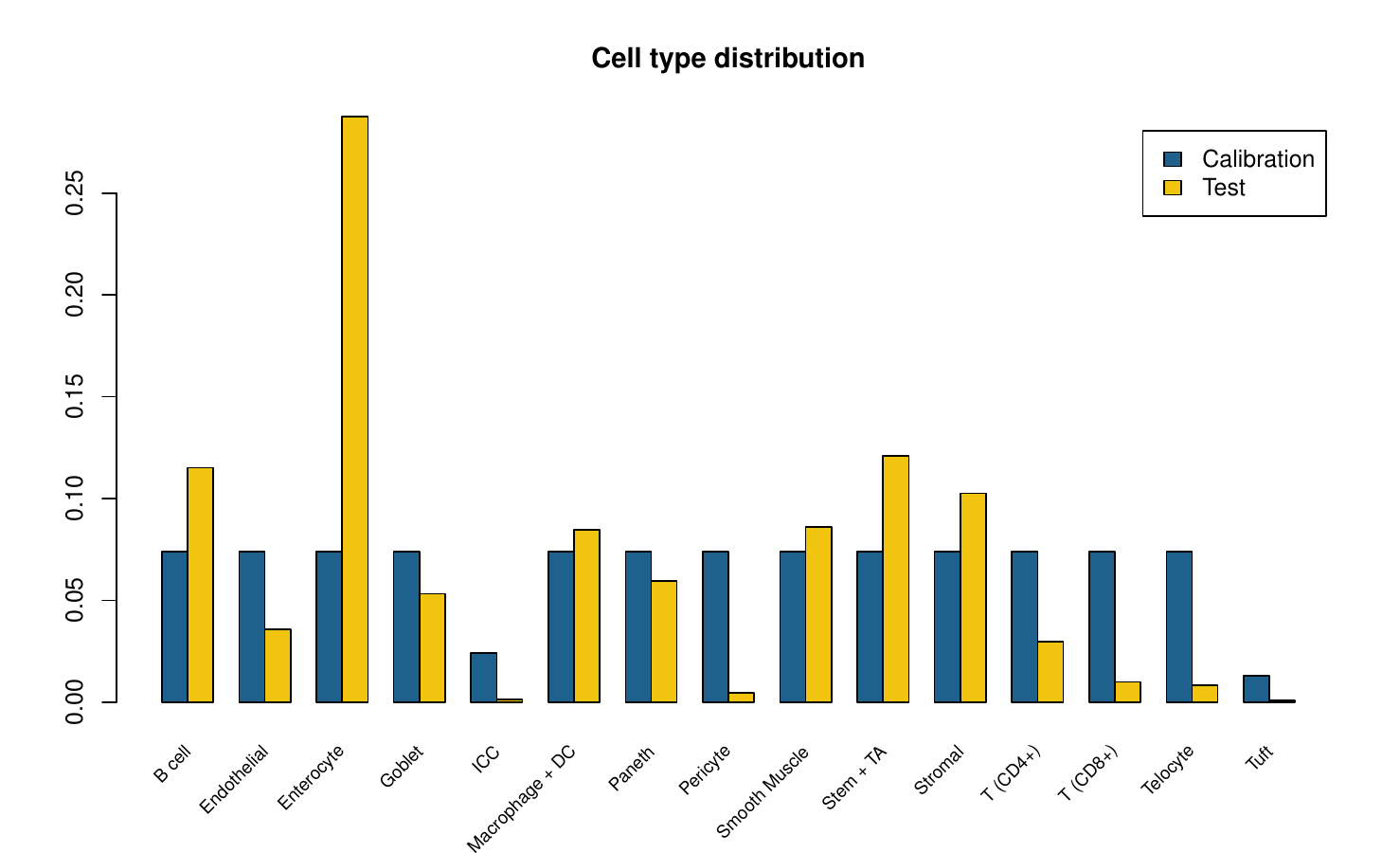}
    \caption{Distribution of cell types in calibration and test dataset for the second case study.}
    \label{fig:dis_over}
\end{figure}

In this second example, we consider a balanced calibration set where each cell type is represented in equal proportions, contrasting with an unbalanced test set. Supplementary Fig. \ref{fig:dis_over} illustrates the distribution of cell types in both datasets.

In this scenario, both the non-corrected conformal and graph-based procedures yield prediction sets with empirical coverages exceeding the nominal coverage. Specifically, they achieve empirical coverages of 0.934 and 0.932, respectively. Despite meeting the nominal coverage criterion (i.e. Equation \eqref{eq:coverage} is satisfied), these sets are too conservative, resulting in reduced statistical power.

To mitigate this issue arising from label shift, we employ our resampling strategy. Following resampling, the empirical coverage decreases to 0.896 for conformal inference and 0.887 for the graph-based procedure.

\begin{figure}
\includegraphics[width=\linewidth]{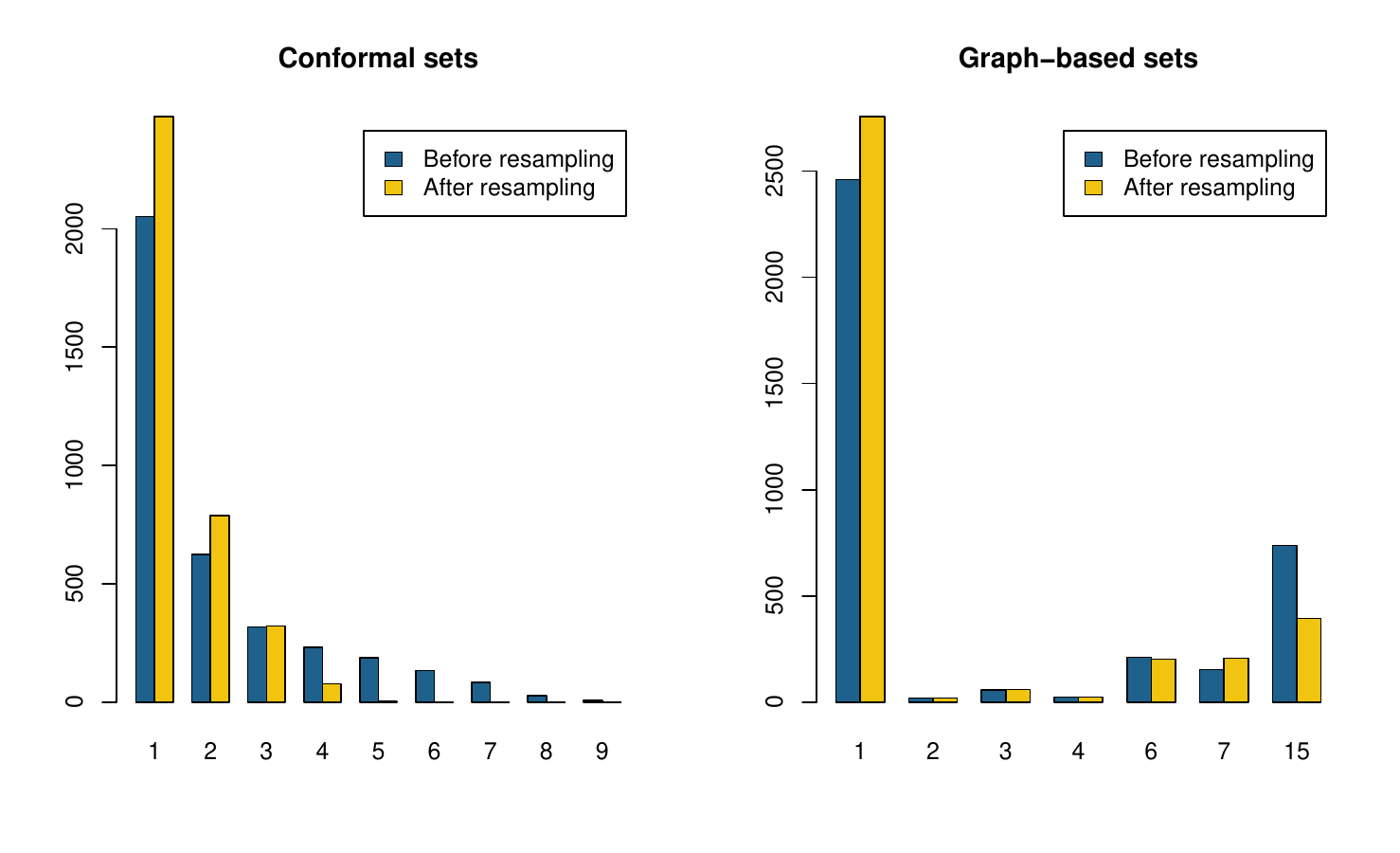}
    \caption{Distribution of size of conformal and graph-based sets before and after resampling.}
    \label{fig:comp_size}
\end{figure}

An insightful comparison concerns the sizes of prediction sets before and after resampling. Supplementary Fig. \ref{fig:comp_size} contrasts the distribution of set sizes obtained without and with the resampling strategy, for both conformal and graph-based approaches. Notably, the maximum size of conformal prediction sets reduces from 9 to 5 after resampling the calibration set. Conversely, without label shift correction, more inconclusive sets (all 15 labels) and fewer single-label sets are observed in graph-based predictions compared to when the resampling strategy is applied.

Finally, results from 100 simulations with different draws of calibration and test sets for this example are summarized in Supplementary Fig. \ref{fig:overEmpCov}. Across these simulations, our approach consistently aligns the empirical distribution to the theoretical one, demonstrating its effectiveness in addressing label shift.
\begin{figure}[H]
\includegraphics[width=\linewidth]{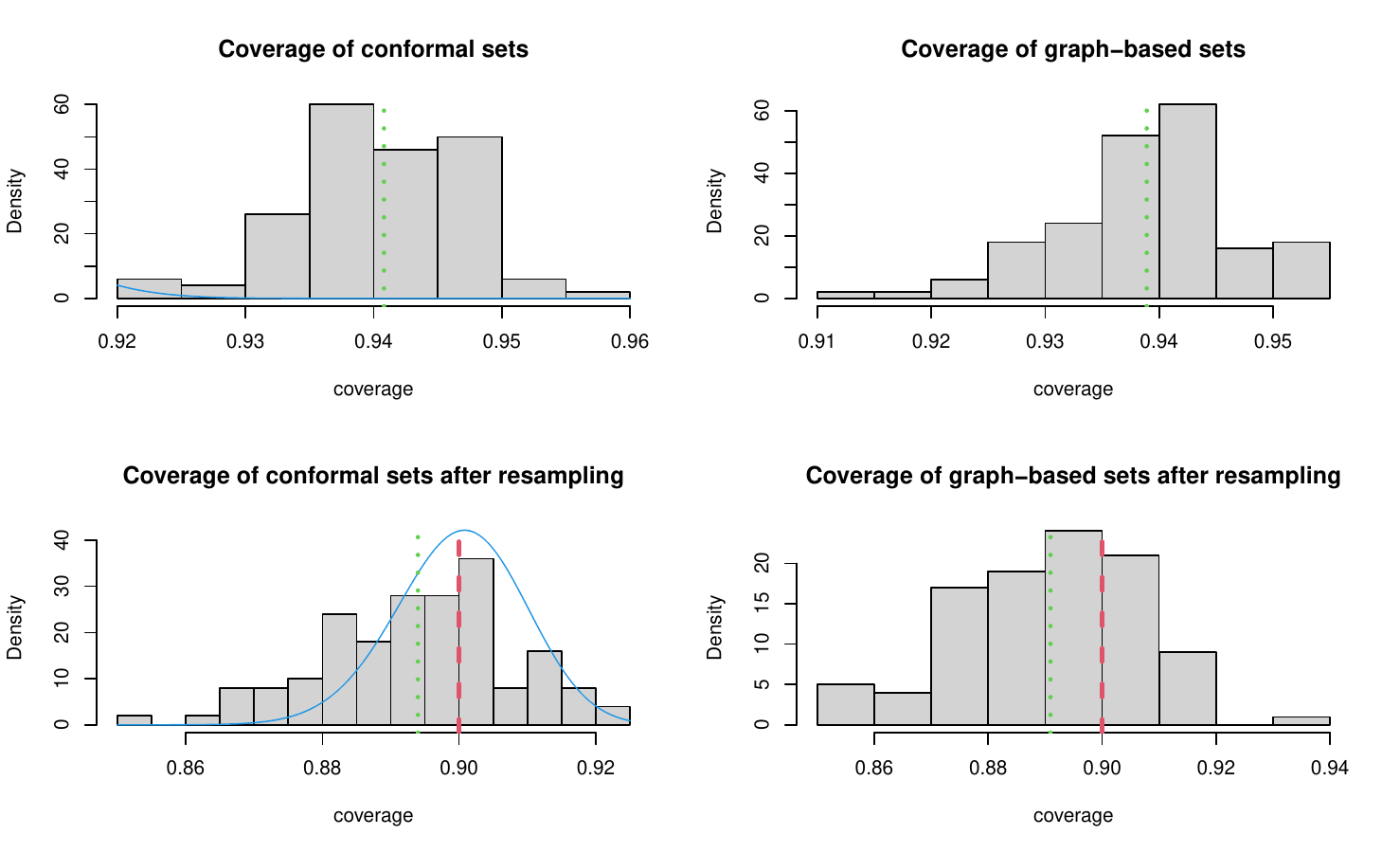}
    \caption{Empirical coverage of the prediction sets over 100 simulations obtained by conformal inference (first column) and by the graph-based procedure (second column) for the second case study. Results in the first row have been obtained without resampling, results in the second row have been obtained resampling. The red dashed lines correspond to the desired coverage, the green dotted ones to the empirical coverage.}
    \label{fig:overEmpCov}
\end{figure}

\end{appendices}

\end{document}